\documentclass[aps,showpacs,onecolumn,12pt]{revtex4}
\usepackage{graphicx}
\usepackage{dcolumn}
\usepackage{bm}

\begin{document}
\title{Large scale shell model calculations for even-even $^{62-66}$Fe isotopes}
\author{P.C.~Srivastava$^{1,2}$, I.~Mehrotra$^1$}
\affiliation{$^1$Nuclear Physics Group, Department of Physics, University of Allahabad, India}
\affiliation{$^2$Grand Acc\'el\'erateur National d'Ions Lourds,
CEA/DSM--CNRS/IN2P3, BP~55027, F-14076 Caen Cedex 5, France}
\date{\today}

\begin{abstract}
 The recently measured experimental data of Legnaro National Laboratories on neutron rich even isotopes of $^{62-66}$Fe with A=62,64,66 have been interpreted in the framework of large scale shell model. Calculations have been performed with a newly derived effective interaction GXPF1A in full $\it{fp}$ space without truncation. The experimental data is very well explained for $^{62}$Fe, satisfactorily reproduced for $^{64}$Fe and poorly fitted for $^{66}$Fe. The increasing collectivity reflected in experimental data when approaching N=40 is not reproduced in calculated values. This indicates that whereas the considered valence space is adequate for $^{62}$Fe, inclusion of higher orbits from $\it{sdg}$ shell is required for describing $^{66}$Fe.    
\end{abstract}

\maketitle

\section{Introduction}
\label{s_intro}
~~The neutron rich nuclei in the $\it{fp}$ shell region are at the focus of attention of the nuclear physics community at present. Unstable nuclei in this region exhibit many new phenomenon such as appearance of new magic numbers and disappearance of well established ones,softening of core at N=28, interplay of collective and single particle properties $\it{etc}$. Neutron rich $\it{fp}$ shell nuclei are also of special interest in astrophysics such as the electron capture rate in supernovae explosion. A large number of neutron rich nuclei can be populated by means of binary reactions such as multinucleon transfer and deep inelastic collisions with stable beam. Such reactions combined with modern $\gamma$ detector arrays have increased substantially the available data on nuclei far from stability. Experimental data on the excited states of neutron rich  unstable isotopes of Ca, Ti and Cr has been made available in recent past \cite{NNDC}. Recently at Legnaro National Laboratories neutron rich Fe isotopes from A=61-66 were populated through multineutron transfer reaction by bombarding a $^{238}$U target with 400MeV $^{64}$Ni beam \cite{Lurandi07}. The identification of $\gamma$-rays belonging to each nucleus was carried out with high precision by coupling the clover detector of Euroball( CLARA) to PRISMA   magnetic spectrometer. This experiment has provided data on level structure of neutron rich Fe isotopes from A=61 to 66.

        ~~~~~~~~~~~~~~~ Nuclear shell model is one of the most powerful tools for giving a quantitative interpretation to the experimental data. The two main ingredients of any shell model calculations are the N-N interaction and the configuration space for valence particles. In principle one can either perform shell model calculations with realistic N-N interaction in unlimited configuration space or with renormalized effective interaction in a limited configuration space. Various methods have been adopted in the literature to obtain effective interaction from realistic N-N interaction. One of the methods is to modify the matrix elements of the   microscopic interaction by carrying an empirical fit to a sufficiently large body of experimental data. Using this approach Honma $\it{et~ al.}$ \cite{Honma04} have derived an effective interaction, GXPF1, from Bonn-C potential for $\it{fp}$ shell and have used it in the shell model calculations in full $\it{fp}$ configuration to obtain the systematics of neutron rich isotopes of Ca, Cr and Ti. Their analysis showed significant deviation from the experimental data towards the end of $\it{fp}$ shell. As GXPF1 was derived by fitting the experimental data of stable isotopes,some modifications in the interaction were required for its use for predicting the structure of unstable nuclei in the $\it{fp}$ shell. In view of this the interaction was modified by changing five matrix elements (three of pairing interaction and two of quadrupole-quadrupole interaction). This modified interaction, known as GXPF1A  \cite{Honma05}, improved the agreement with experimental data for unstable isotopes in $\it{fp}$ region. Lunardi $\it{et~ al.}$ \cite{Lurandi07} have interpreted the results of their experiments on $^{61-66}$Fe isotopes by performing large scale shell model calculations with an effective interaction `fpg' described in Ref \cite{Sorlin02}. In their calculation an inert core of $^{48}$Ca is considered and the valence space chosen is the whole $\it{fp}$ shell for the protons and the p$_{3/2}$ f$_{5/2}$  p$_{1/2}$ and g$_{9/2}$ orbitals for the neutrons.  In order to reduce the dimensions of the matrices involved a truncation was put on the number of particles allowed to be excited: a maximum of x protons were allowed to be excited from the f$_{7/2}$ to the rest of the $\it{fp}$  shell and (n-x) neutrons from three $\it{fp}$  orbitals to g$_{9/2}$ orbital where n=5.

          In the present work we have performed large scale shell model calculations on neutron rich $^{62-66}$Fe isotopes with GXPF1A interaction without any truncation. The calculations have been carried out in valence space of full  $\it{fp}$ shell consisting of 0f$_{7/2}$ 1p$_{3/2}$ 0f$_{5/2}$ 1p$_{1/2}$ orbitals and treating $^{40}_{20}$Ca as the inert core. No restriction has been put on the number of particles which can be excited to higher level.The aim of this paper is to test the suitability of GXPF1A interaction towards the end of $\it{fp}$  shell. Also by comparing the results of present work with that of Ref \cite{Lurandi07} one can get some insight on the role of g$_{9/2}$ orbit in explaining the data.

 The paper is organized as follows: Section II gives details of the calculation. Results and discussion are given in Section III. Finally in section IV conclusions are given.

\section{ Details of Calculation}

\subsection{Configuration space}
Large scale shell model calculations have been performed for neutron rich even Fe isotopes with A=62,64,66 treating $^{40}$Ca as inert core. The configuration space for valence particles is taken as full $\it{fp}$ shell which is made up of all Pauli allowed combinations of valence particles in the  0f$_{7/2}$ 1p$_{3/2}$ 0f$_{5/2}$ 1p$_{1/2}$ orbitals for both protons and neutrons. The single particle energies for model space 0f$_{7/2}$, 1p$_{3/2}$,  1p$_{1/2}$, 0f$_{5/2}$ are -8.6240, -5.6793, -4.1370 and -1.3829 MeV respectively.
\subsection{Effective Interaction}
 The calculations have been performed with a newly derived effective interaction GXPF1A obtained from a fit to the experimental data of unstable nuclei in the $\it{pf}$ shell. Honma $\it{et~al.}$ \cite{Honma05} initially derived an effective interaction, GXPF1, starting from Bonn-C  potential by modifying 70 well determined combinations of 4 single particle energies and 195 two body matrix elements by iterative fitting calculations to about 699 experimental energy data out of 87 stable nuclei. These authors have tested the GXPF1 interaction for the shell model calculations in the full $\it{fp}$ shell extensively \cite{Honma04} from various viewpoints such as binding energies, electromagnetic moments and transitions, and excitation spectra in the wide range of $\it{fp}$ shell nuclei. They observed that the deviation of the shell model prediction from available experimental data appeared to be  sizable in binding energies of N $\geq$ 35 nuclei and in magnetic moments of Z $\geq$ 32 even-even nuclei. For the unstable nuclei, whereas the experimental data on Ca and Cr isotopes were well explained with GXPF1A, the experimental value of the first excited 2$^+$ state of $^{56}$Ti was lower than the predicted value by about 0.4MeV requiring modification of the interaction. The interaction was modified by these authors by changing 5 two body matrix elements in the $\it{fp}$ shell: 3 pairing interaction matrix elements were made slightly weaker and two quadrupole-quadrupole matrix elements were made slightly stronger \cite{Honma05}. The modified interaction, referred to as GXPF1A, gave improved description simultaneously for all these three isotope chains and is reliable for use in shell model calculations to explain the data on unstable nuclei.
\subsection{Computer code}
  The calculations have been performed  at the SGI-cluster computer at GANIL with the code {\tt ANTOINE }~\cite{Caurier89,Caurier99}. In this code the problem of giant matrices is solved by splitting the valence space into two parts, one for proton and another for the neutron. The states of the basis are written as the product of two Slater determinants (SD), one for protons and another for neutrons: $\mid I\rangle= \mid i,\alpha $. (Here capital letter refers to full space and lower case letters refer to subspaces of proton and neutron).
The Slater determinants i and $\alpha$ can be classified by their $\it{M}$ values,$\it{M_1}$ and $\it{M_2}$. The total $\it{M}$ being fixed, the SD of the two subspaces will be associated only if $\it{M_1+M_2= M}$. Dimensions of the matrices involved for $^{62-66}$Fe in m-scheme are given in Table 1 for different J states.
\begin{table}[h]
\begin{center}
\caption{
Dimensions of matrices involved for  $^{62-66}$Fe  in $\it{m}$ scheme for $\it{fp}$ shell .}
\label{t_dim}
\begin{tabular}{rcrc}
\hline
\hline
 J$^\pi$ &~~ $^{62}$Fe &~~ $^{64}$Fe  & ~~ $^{64}$Fe \\		   
\hline
 0$^+$ &~~ 14625240 &~~ 634744 &~~ 3952  \\
 2$^+$ &~~ 14358186 &~~ 620567 &~~ 3815  \\
 4$^+$ &~~ 13586777 &~~ 580340 &~~ 3496  \\
 6$^+$ &~~ 12386760 &~~ 518387 &~~ 2984  \\
 8$^+$ &~~ 10876432 &~~ 442325 &~~ 2406 \\ 
\hline  
\hline            
\end{tabular}
\end{center}
\end{table}  
\begin{figure}[h]
\begin{center}
\includegraphics[width=9cm]{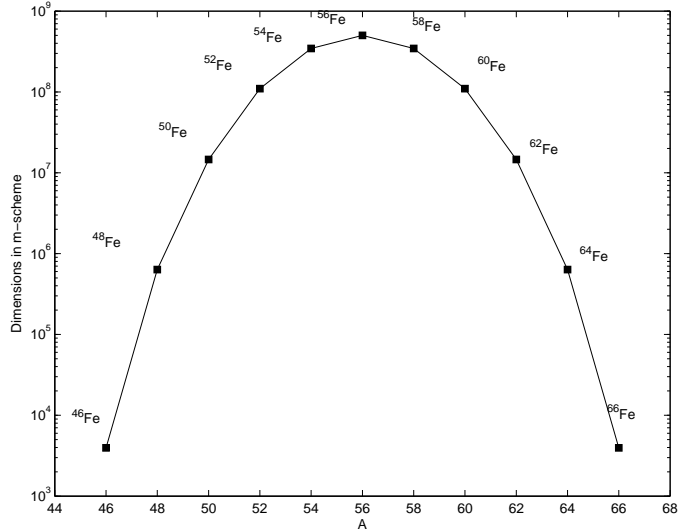}
\caption{
Dimensions of the matrices for 0$^+$ state in m-scheme as a function of A.}
\label{f_fed}
\end{center}
\end{figure}

  The dimensionality of the matrices is maximum at midshell and decreases very fast towards the beginning and at the end of the shell as shown in Fig.1. The computing time for $^{62}$Fe  was three days compared to two hours for $^{66}$Fe.
 
\section{ Results and Discussions}

 \subsection{Excitation energies and comparison with the data}

The calculated energy levels obtained with GXPF1A interaction for even Fe isotopes are shown in Fig. 2-4 and compared with the experimental data and also with the results obtained with `$\it {fpg}$'  interaction in a truncated configuration space. It is observed that large scale shell model calculation with GXPF1A interaction in full $\it{fp}$ space gives very good agreement with the experimental data for $^{62}$Fe. For 2$^+$ , 4$^+$  and 6$^+$  states of $^{62}$Fe the discrepancies with experimental data are 70, 5 and 138 KeV respectively whereas with `$\it {fpg}$' interaction the corresponding values are 6, 219 and 494 KeV. But the discrepancy with the experimental data increases as we go towards N=40. `$\it{fpg}$' interaction with truncation on the number of particles getting excited gives better result. This shows the importance of g$_{9/2}$ orbital in explaining the data for A=66. This can be understood in terms of decrease in the energy gap between $\it{fp}$ shell and 1g$_{9/2}$ orbital in going towards N=40 and can be attributed to the proton neutron monopole tensor interaction \cite{Honma04}. Variation of first 2$^+$ excited state energy levels E(2$^+$) with neutron number is shown in Fig.5. Thus the recent data on $^{62}$Fe can be explained very well in the frame work of large scale shell model calculations with GXPF1A interaction in full $\it{fp}$ space with no truncation, without including 0g$_{9/2}$ orbital. The fit is reasonably satisfactory for $^{64}$Fe, but agreement with the experimental data on $^{66}$Fe is not good.
\begin{figure}[h]
\begin{center} 
\includegraphics[width=9cm]{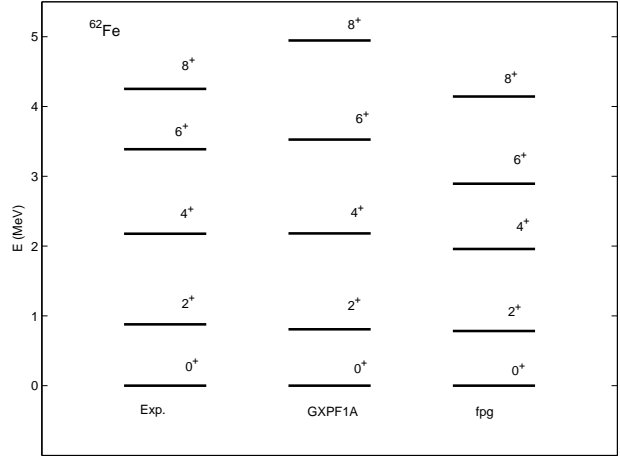}
\caption{
Calculated energy levels of $^{62}$Fe with GXPF1A interaction compared with the experimental data and the previous theoretical work using `$\it{fpg}$' interaction with truncation from Ref. \cite{Lurandi07}.}
\label{f_fe62}
\end{center}
\end{figure}
\begin{figure}[h]
\begin{center}
\includegraphics[width=9cm]{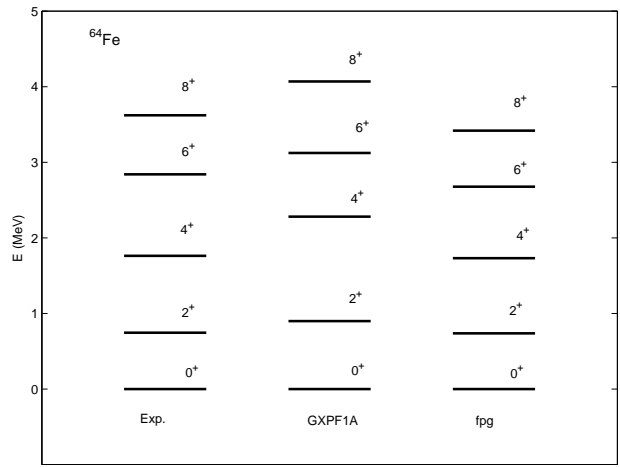}
\caption{
Same as Fig.2 for $^{64}$Fe.}
\label{f_fe64}
\end{center}
\end{figure}

\begin{figure}
\begin{center}
\includegraphics[width=9cm]{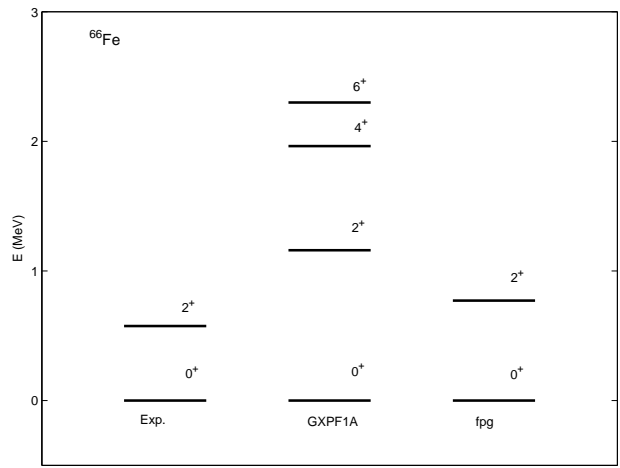}
\caption{
Same as Fig.2 for $^{66}$Fe.}
\label{f_fe66}
\end{center}
\end{figure}

\begin{figure}[h]
\begin{center}
\includegraphics[width=9cm]{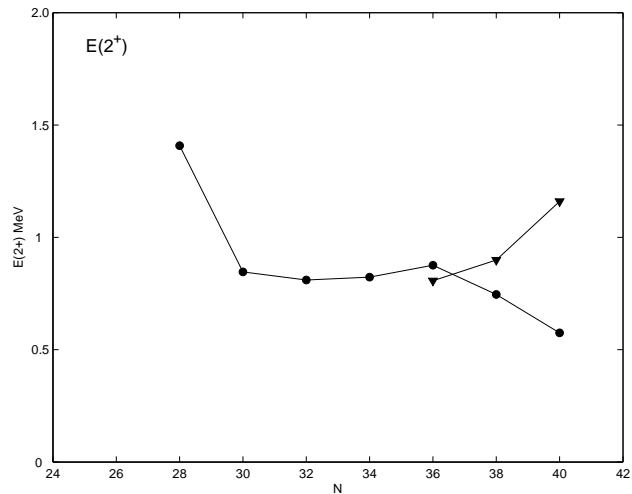}
\caption{
First 2$^+$ energy levels as a function of neutron number $\it{N}$. Experimental data are shown by filled circle and results from present work by filled triangle.}
\label{f_feE2}
\end{center}
\end{figure} 
\pagebreak
 \subsection{Wave functions}
The wave functions for  $^{62-66}$Fe isotopes are shown in table II. The most dominant contribution in the ground state of  $^{62}$Fe is (0f$_{7/2})^8$, (1p$_{3/2})^4$,(1p$_{1/2})^0$, (0f$_{5/2})^4$ whereas that of $^{64}$Fe is (0f$_{7/2})^8$, (1p$_{3/2})^4$, (1p$_{1/2})^2$, (0f$_{5/2})^4$ indicating change in the ordering of 1p$_{1/2}$ and 0f$_{5/2}$ levels.

\begin{table}
\begin{center}
\caption{
Main configurations in the wave functions of the ground state and first exited state for $^{62-66}$Fe calculated with GXPF1A interaction.}
\label{t_con}
\begin{tabular}{rcrcr}
\hline
\hline
 Nuclei  & J$^\pi$ &~~~~~~Wave function & & Probability \\
           ~&~  & Neutron & Proton  & ~ \\		   
\hline
 $^{62}$Fe   &~ 0$_{gs}^+$&~ (0f$_{7/2})^8$,(1p$_{3/2})^4$,(1p$_{1/2})^0$,(0f$_{5/2})^4$ &~~(0f$_{7/2})^6$ & 34.4  \\
    &~ 2$_{1}^+$&~ (0f$_{7/2})^8$,(1p$_{3/2})^4$,(1p$_{1/2})^0$,(0f$_{5/2})^4$ &~~(0f$_{7/2})^6$ & 28.8 \\
 $^{64}$Fe  &~ 0$_{gs}^+$ &~ (0f$_{7/2})^8$,(1p$_{3/2})^4$,(1p$_{1/2})^2$,(0f$_{5/2})^4$ &~~(0f$_{7/2})^6$ & 73.3  \\
    & ~2$_{1}^+$&~ (0f$_{7/2})^8$,(1p$_{3/2})^4$,(1p$_{1/2})^2$,(0f$_{5/2})^4$ &~~(0f$_{7/2})^6$ & 71.5 \\
 $^{66}$Fe & 0$_{gs}^+$ &~ (0f$_{7/2})^8$,(1p$_{3/2})^4$,(1p$_{1/2})^2$,(0f$_{5/2})^6$ &~~(0f$_{7/2})^6$ &  96.6 \\
    & ~2$_{1}^+$&~ (0f$_{7/2})^8$,(1p$_{3/2})^4$,(1p$_{1/2})^2$,(0f$_{5/2})^6$ &~~(0f$_{7/2})^6$ & 95.5 \\ 
\hline  
\hline            
\end{tabular}
\end{center}
\end{table}

   \subsection{Electromagnetic properties}
The calculated B(E2) values for 2$_1^+$ $\rightarrow$ 0$_{ gs}$ transition are shown in table 3 for $^{62-66}$Fe, using standard effective charges e$_{\pi}$=0.5 and e$_{\nu}$=1.5 respectively \cite{Bohr,Brown05}. A free neutron does not have any charge but nucleons in the nucleus can polarize the core by interacting with nucleons of the core. This is reflected by giving the neutron an effective charge. The B(E2) values obtained in the present calculation are plotted in Fig.6 as a function of neutron number along with the available experimental data for lighter Fe isotopes in the same figure. 

         ~~~~~~The results obtained with GXPF1A show a large increase in the value of E$_{ex}(2^+)$ and decrease in BE(2$^+$) at N=40 - a feature typical of shell closer at N=40. Similar pattern is obtained for N=28. In contrast to this the experimental value of first 2$^+$ exited state decreases with increasing number of neutrons when approaching N=40. The drop in the excitation energy is a signature for increasing collectivity. This feature could not be explained with GXPF1A and the calculated excitation energy of 2$^+$ state is predicted too high. This indicates that the considered valence space is not large enough to account for increase of collectivity at N=40. The results obtained with  `$\it {fpg}$' interaction are better. This shows the importance of g$_{9/2}$ orbital in explaining the data for A=66. 
\begin{table}[h]
\begin{center}
\caption{}
\begin{tabular}{rcrc}
\hline
\hline

 -~~~~~~~ &~~ $^{62}$Fe &~~ $^{64}$Fe  &~~  $^{64}$Fe \\		   
\hline
 E(2$^+$)(MeV) &~~ 0.807 &~~ 0.899 &~~ 1.160  \\
 E(4$^+$)(MeV) &~~ 2.181 &~~ 2.281 &~~ 1.963  \\
 Q(2$^+$)(efm$^2$) &~~ -26 &~~ -19 &~~ -17  \\
 B(E2)(W.u.)&~~ 14.8 &~~ 10.1 &~~ 5.4  \\
 
\hline  
\hline            
\end{tabular}
\end{center}
\end{table}

\begin{figure}[h]
\begin{center}
\includegraphics[width=9cm]{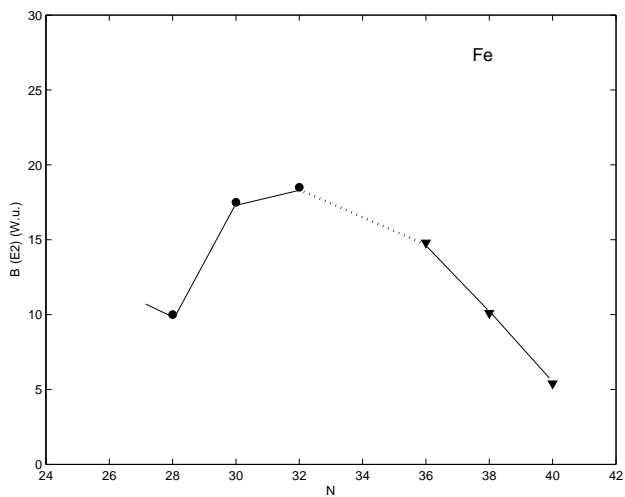}
\caption{ 
B(E2;2$_1^+$$\rightarrow$ 0$^+$) for $^{54-58}$Fe from Ref \cite{Honma04} by filled circle and  results from present work by filled triangle for $^{62-66}$Fe .}
\label{f_bE2}
\end{center}
\end{figure}

\section{ Conclusions}
 
In the present work large scale shell model calculations have been performed for neutron rich even isotopes of Fe with A=62,64,66 in full $\it{fp}$ space
without truncation with recently derived GXPF1A interaction suitable for use in $\it{fp}$ shell for unstable nuclei. The experimental data is very well reproduced for $^{62}$Fe and the agreement is better than the earlier calculations carried out in  $\it{fpg}$ configuration space with truncation imposed on the maximum number of particles getting excited to higher levels. This indicates that full  $\it{fp}$ space is sufficient for $^{62}$Fe nucleus. The agreement with the experimental data is reasonably satisfactory for $^{64}$Fe and is almost same as those obtained with $\it{fpg}$ configuration space. The agreement gets worse for $^{66}$Fe showing the inadequacy of the chosen configuration space.

{\bf Acknowledgement}\\
 The first author would like to thank P.~Van Isacker, M.~Rejumund, E.~Caurier and F.~Nowacki for providing access to the shell model code ANTOINE at GANIL. This work was financially supported by the Sandwich PhD programme of the Embassy of France in India.

\end{document}